\documentclass[conference]{IEEEtran}
\usepackage{cite}
\usepackage{amsmath,amssymb,amsfonts}
\usepackage{algorithmic}
\usepackage{graphicx}
\usepackage{comment}
\usepackage{textcomp}
\usepackage[dvipsnames]{xcolor}
\usepackage{pgfplots}
\pgfplotsset{compat = newest}
\usepackage{listings}
\usepackage{tikz}
\usepackage[inkscapeformat=png]{svg}
\usetikzlibrary{fit, shapes.geometric, arrows.meta, positioning, calc}
\usepackage{booktabs} 
\usepackage{longtable} 
\usepackage{array} 
\usepackage{url}
\usepackage{titlesec}

\def\BibTeX{{\rm B\kern-.05em{\sc i\kern-.025em b}\kern-.08em
    T\kern-.1667em\lower.7ex\hbox{E}\kern-.125emX}}
\begin{document}

\lstset{
  basicstyle=\ttfamily,
  columns=fullflexible,
  breaklines=true,
  postbreak=\mbox{\textcolor{red}{$\hookrightarrow$}\space},
  frame=single,
  language=Verilog,
  showstringspaces=false,
  commentstyle=\color{gray},
  keywordstyle=\color{blue},
  framesep=10pt,
  breakindent=0pt,
  columns=fullflexible,
  xleftmargin=10pt,xrightmargin=10pt
}

\definecolor{armygreen}{rgb}{0.29, 0.33, 0.13}
\definecolor{cadmiumgreen}{rgb}{0.0, 0.42, 0.24}
\definecolor{darkgreen}{rgb}{0.0, 0.2, 0.13}
\definecolor{darkolivegreen}{rgb}{0.33, 0.42, 0.18}
\definecolor{emerald}{rgb}{0.25, 0.75, 0.42}
\definecolor{light-gray}{gray}{0.80}
\definecolor{dark-gray}{gray}{0.5}

\newcommand\akturk[1]{\textcolor{emerald}{#1}}
\newcommand\selim[1]{\textcolor{blue}{#1}}
\newcommand\todo[1]{\textcolor{red}{TODO: #1}}
\newcommand\note[1]{\textcolor{red}{NOTE: #1}}
\newcommand\question[1]{\textcolor{violet}{Q: #1}}

\newcommand{\HighlightA}{\makebox[0pt][l]{\color{light-gray}\rule[-2pt]{0.9\linewidth}{9pt}}}
\newcommand{\HighlightB}{\makebox[0pt][l]{\color{emerald}\rule[-2pt]{0.9\linewidth}{9pt}}}
\newcommand{\HighlightC}{\makebox[0pt][l]{\color{light-gray}\rule[-2pt]{1\linewidth}{9pt}}}
\newcommand{\HighlightD}{\makebox[0pt][l]{\color{light-gray}\rule[-2pt]{0.938\linewidth}{9pt}}}

\title{Zero-Shot RTL Code Generation with Attention Sink Augmented Large Language Models\\
}

\author{\IEEEauthorblockN{Selim Sandal}
\IEEEauthorblockA{\textit{Department of Computer Science} \\
\textit{Ozyegin University}\\
Istanbul, Turkey \\
selim.sandal@ozu.edu.tr} 
\and
\IEEEauthorblockN{Ismail Akturk}
\IEEEauthorblockA{\textit{Department of Computer Science} \\
\textit{Ozyegin University}\\
Istanbul, Turkey \\
ismail.akturk@ozyegin.edu.tr} 
}

\maketitle

\begin{abstract}

The design and optimization of hardware have traditionally been resource-intensive, demanding considerable expertise and dependence on established design automation tools. This paper discusses the possibility of exploiting large language models to streamline the code generation process in hardware design. In contrast to earlier studies, this paper aims to use large language models that accepts high-level design specifications through a \textit{single prompt} to generate corresponding Register-Transfer Level (RTL) code. The ability to use large language models on RTL code generation not only expedites design iteration cycles but also facilitates the exploration of design spaces that have computational challenges for conventional techniques. Through our evaluation, we demonstrate the shortcoming of existing attention mechanisms, and present the abilities of language models to produce functional, optimized, and industry-standard compliant RTL code when a novel attention mechanism is used. These findings underscore the expanding role of large language models in shaping the future landscape of architectural exploration and automation in hardware design.

\end{abstract}


\section{Introduction}


The advanced manufacturing technology has led to the creation of increasingly complex hardware architectures, propelling progress in various sectors including healthcare, telecommunications, and autonomous systems. However, the process of designing and optimizing these architectures remains cumbersome, inefficient, and highly dependent on human expertise. Current methodologies necessitate specialized skills in hardware description languages, such as Verilog or VHDL, and a thorough understanding of design automation tools. This complexity not only results in labor intensive design processes, but also poses limitations on exploring the vast and intricate design spaces. As a result, the industry has been struggling with higher costs and prolonged development cycles.

Recently, the application of machine learning techniques, specifically large language models (LLMs), has been explored \cite{blocklove2023chipchat,thakur2022benchmarking,chang2023chipgpt,thakur2023verigen,fu2023gpt4aigchip} as a solution to automate and optimize code generation tasks. These models show potential in generating module-level code in industry standards. However, their capabilities have been largely confined to writing single, isolated modules, leaving a gap in the full-scale architectural exploration and design automation.

In the context of LLMs, \textit{zero-shot} denotes the model's capacity to perform tasks for which it has not explicitly trained on. In hardware design, this implies the model's ability to interpret high-level design specifications even in the absence of prior examples related to similar tasks. This approach is particularly valuable, as it facilitates adaptable and responsive handling of a diverse array of design requirements, empowering designers to explore innovative solutions without the need for the model to undergo specialized training for each distinct type of hardware design task. Note that while a given LLM may have been trained on a corpus that includes RTL code (e.g., Verilog), along with other non-hardware-related information, it has not undergone direct training to translate high-level design concepts into RTL. This distinction is crucial for appreciating the contributions of this work.

Currently, existing LLMs struggle generating large code bases that align seamlessly with the provided prompt, primarily attributed to the limitations imposed by the context window size. This work aims to demonstrate the feasibility of zero-shot prompting to generate high quality RTL code. To do so, we present an approach that leverages the power of LLMs in conjunction with a recently introduced attention mechanism called \textit{attention sink}~\cite{xiao2023streamingllm}. In this scenario, a singular prompt encapsulates high-level design specifications, and the underlying language model, enhanced with attention sink, is employed to generate the comprehensive RTL code. This not only significantly expedites the design process, but also enables tackling design spaces that have hitherto been computationally infeasible for traditional methods.

Below is the summary of main contributions and novelty of this work:

\begin{itemize}
    \item survey the shortcoming of existing LLMs with available attention mechanisms in RTL code generation;
    \item present a methodology that leverages LLMs for automated RTL code generation from high-level specifications, without the need for model fine-tuning specific to RTL (note that fine-tuning for RTL is challenging due to lack of industry-standard compliant comprehensive training data);
    \item integration of a newly introduced attention mechanism, attention sink, with LLMs that addresses the challenges posed by traditional LLMs in managing long-sequence tasks;
    \item  demonstrate LLMs ability (with attention sink mechanism)  to interpret and generate functional, optimized, and industry-standard compliant RTL code for an Neural Processing Unit (NPU);
    \item provide insights for future innovations in LLM-based hardware design automation tools.
\end{itemize}


In the following, we provide some background on existing efforts within the domain and elucidate their limitations. Subsequently, we provide the specifications of the NPU used as a case study. Thereafter, we outline our proposed approach, explain the methodology employed in evaluation, and present our findings. Our findings suggest that LLMs that enhanced with the novel attention mechanism can produce functional, optimized, and industry-standard compliant RTL code. This work contributes to the ongoing discussions about the potential of LLMs in facilitating more efficient  design automation and architectural exploration.

\section{Background}
\subsection{Existing LLMs and their limitations in RTL code generation}
\label{subsec:llm-limitations}



The design of emerging hardware architectures (e.g., AI accelerators) involves complex logic circuits and sophisticated elements tailored to specific tasks, 
often demanding prompts with long and detailed specifications. 
To generate effective RTL code, the prompts must encompass comprehensive functional requirements, specifying the design's operational range, data processing capabilities, and logical structures for the desired computation. 
When utilizing large language models for RTL code generation, accurate interpretation and implementation of these detailed specifications are vital for creating functionally effective and technically accurate hardware designs. Nevertheless, we note that existing LLMs often produce RTL code that is either non-synthesizable or functionally incorrect. Below, we outline the most prevalent failures observed in LLMs when generating RTL code.

\noindent {\bf Redundant copies of variables}:
LLMs often \textit{forget} the original prompt (mainly due to the limited context window) and regenerate the variables from the other modules, instead of instantiating these modules, as shown in Listing~\ref{listing-unneccessary-copy}. 

\noindent {\bf Hallucinating Variables and Specifications:}
A notable challenge in LLM generated RTL code is \textit{hallucination}, where the model diverges from providing the requested code and instead generates extraneous variables or specifications. 
In LLMs, the Key-Value (KV) cache plays a crucial role in the transformer attention mechanism~\cite{vaswani2023attention}, a core component that determines how the model processes and generates text. The KV cache is used to store and manage 'keys' and 'values,' which are elements that represent aspects of the input tokens and their associated contextual information. 

In LLMs, using attention mechanisms like dense and windowed attention reveals context window limitations, especially when processing text longer than the KV cache size. Dense attention may cause model to hallucinate, especially with variable definitions, as shown in Listing~\ref{listing-repeatation}, due to a weakened context mechanism that loses earlier code segments. While windowed attention is more memory efficient, it can increase language modeling perplexity, thereby reducing the model's coherence and accuracy in code generation.

\begin{lstlisting}[language={Verilog}, caption={Redundant Copies of Variables}, basicstyle=\small, label=listing-unneccessary-copy, escapechar=!, float=tp, belowskip=-0.8 \baselineskip]
module bfloat16_addsub #(
   parameter WIDTH = 16
) (
   input logic [WIDTH-1:0] a,
   input logic [WIDTH-1:0] b,
   output logic [WIDTH-1:0] result,
   input is_sub
);
   logic [WIDTH-1:0] a_sign, b_sign, result_sign;
   logic [WIDTH-1:0] a_exp, b_exp, result_exp;
   logic [WIDTH-1:0] a_mant, b_mant, result_mant;
   !\HighlightA{}!logic [WIDTH-1:0] shift_amount; 
   !\HighlightA{}!logic [WIDTH-1:0] shifted_mant_a, shifted_mant_b;
   !\HighlightA{}!logic [WIDTH-1:0] sum;                           
   !\HighlightA{}!logic [WIDTH-1:0] carry;                     
   
module bfloat16_div #(
   parameter WIDTH = 16
) (
   input logic [WIDTH-1:0] a,
   input logic [WIDTH-1:0] b,
   output logic [WIDTH-1:0] result
);
   !\HighlightB{}!logic [WIDTH-1:0] a_sign, b_sign, result_sign;     
   !\HighlightB{}!logic [WIDTH-1:0] a_exp, b_exp, result_exp;      
   !\HighlightB{}!logic [WIDTH-1:0] a_mant, b_mant, result_mant;   
   !\HighlightB{}!logic [WIDTH-1:0] quotient;                      
   !\HighlightB{}!logic [WIDTH-1:0] leading_zeros;

module processing_element #(
   parameter WIDTH = 16
) (
   input logic clk,
   input rst,
   input [3:0] opcode,
   input [WIDTH-1:0] a_in,
   input [WIDTH-1:0] b_in,
   output logic [WIDTH-1:0] result
);
   !\HighlightB{}!logic [WIDTH-1:0] a_sign, b_sign, result_sign;   
   !\HighlightB{}!logic [WIDTH-1:0] a_exp, b_exp, result_exp;      
   !\HighlightB{}!logic [WIDTH-1:0] a_mant, b_mant, result_mant; 
   !\HighlightA{}!logic [WIDTH-1:0] shift_amount;                
   !\HighlightA{}!logic [WIDTH-1:0] shifted_mant_a, shifted_mant_b;
   !\HighlightA{}!logic [WIDTH-1:0] sum;                           
   !\HighlightA{}!logic [WIDTH-1:0] carry;                         
   logic [WIDTH-1:0] product;
   !\HighlightB{}!logic [WIDTH-1:0] quotient;                      
\end{lstlisting}


Similarly, Listing~\ref{listing-hallucination-problem} shows that instead of producing the desired code, LLM continues to elaborate or invent specifications that were not part of the original prompt. 
This behavior can be attributed to the model's difficulty in contextual understanding and its tendency to fill gaps in its knowledge by generating plausible-sounding, but ultimately useless content. 
Hallucination in code generation is particularly problematic as it can lead to bloated, convoluted designs that stray significantly from the intended outcome, necessitating rigorous review and correction by human experts. Addressing this challenge requires refining the model's ability to stay focused on the initial prompt and its relevant context, ensuring that the generated code remains aligned with the specific requirements of the RTL design task.

\begin{lstlisting}[language={Verilog}, caption={Hallucinating Variables}, basicstyle=\small, label=listing-repeatation, escapechar=!, float=tp, belowskip=-0.8 \baselineskip]
logic [15:0] a_sign;
logic [15:0] b_sign;
logic [15:0] result_sign;
logic [15:0] result_exp;
logic [15:0] result_mantissa;
logic [15:0] a_exp;
logic [15:0] b_exp;
logic [15:0] a_mantissa;
logic [15:0] b_mantissa;
logic [15:0] a_mantissa_shifted;
logic [15:0] b_mantissa_shifted;
!\HighlightC{}!logic [15:0] a_mantissa_shifted_sub;  
!\HighlightC{}!logic [15:0] b_mantissa_shifted_sub; 
!\HighlightC{}!logic [15:0] a_mantissa_shifted_add;
!\HighlightC{}!logic [15:0] b_mantissa_shifted_add;
!\HighlightC{}!logic [15:0] a_mantissa_shifted_sub_normalized;
!\HighlightC{}!logic [15:0] b_mantissa_shifted_sub_normalized;
!\HighlightC{}!logic [15:0] a_mantissa_shifted_add_normalized;
!\HighlightC{}!logic [15:0] b_mantissa_shifted_add_normalized;
!\HighlightC{}!logic [15:0] a_mantissa_shifted_sub_normalized_result;
!\HighlightC{}!logic [15:0] b_mantissa_shifted_sub_normalized_result;
!\HighlightC{}!logic [15:0] a_mantissa_shifted_add_normalized_result;
!\HighlightC{}!logic [15:0] b_mantissa_shifted_add_normalized_result;
!\HighlightC{}!logic [15:0] a_mantissa_shifted_sub_normalized_result_final;
!\HighlightC{}!logic [15:0] b_mantissa_shifted_sub_normalized_result_final;
\end{lstlisting}


\begin{lstlisting}[language={}, caption={Hallucinating Specification}, basicstyle=\small, label=listing-hallucination-problem, breaklines=false, escapechar=!]
The control module receives commands and data via a wishbone 
bus from the RISC-V cpu.
The control module decodes the instruction and executes the 
operation.
The control module also manages the vector registers and scalar 
registers.
The control module also manages the processing elements.
The control module also manages the instruction decoder.
The control module also manages the bfloat16 operation 
modules.
The control module also manages the wishbone bus.
The control module also manages the RISC-V cpu.
The control module also manages the NPU.
The control module also manages the bfloat16 operation 
modules.
\end{lstlisting}


\noindent {\bf Corrupt output}: 
Corrupt output in the generated RTL code 
can primarily be attributed to the inherent limitations of window attention mechanism in handling long sequences. 
While this method is effective for shorter texts, it struggles with longer sequences, such as detailed RTL code specifications. In these situations, as the model progresses through the input, crucial tokens from the initial sections might fall outside the window's reach, leading to a loss of context. This results in the model generating code that is either incomplete or nonsensical, as illustrated in Listing~\ref{listing-corrupt-output}.



\begin{lstlisting}[language={Verilog}, caption={Corrupt Output}, basicstyle=\small, label=listing-corrupt-output, escapechar=!, float=tp, belowskip=-0.8 \baselineskip]
module processing_element #(
   parameter WIDTH = 16
) (
   input logic clk,
   input rst,
   input [3:0] opcode,
   input [WIDTH-1:0] a_in,
   input [WIDTH-1:0] b_in,
   output logic [WIDTH-1:0] result
);
   logic [WIDTH-1:0] a_sign, b_sign, result_sign;
   logic [WIDTH-1:0] a_exp, b_exp, result_exp;
   logic [WIDTH-1:0] a_mant, b_mant, result_mant;
   logic [WIDTH-1:0] shift_amount;
   logic [WIDTH-1:0] shifted_mant_a, shifted_mant_b;
   logic [WIDTH-1:0] sum;
   logic [WIDTH-1:0] carry;
   logic [WIDTH-1:0] product;
   logic [WIDTH-1:0] quotient;
!\HighlightC{}!   assign a_sign = a_in[_____________!\HighlightD{}!------------------------
!\HighlightC{}!
!\HighlightC{}!----------------------------------------
!\HighlightC{}!
!\HighlightC{}!--------------------
!\HighlightC{}!---------------------
!\HighlightC{}!-------------
!\HighlightC{}!------
\end{lstlisting}

\noindent {\bf Variable type and size mismatch:}
The variable size or type mismatches arise not from erroneous processing of data per se, but from the context-dependent nature of tokenization in fine-tuned LLMs. Tokenization
involves breaking down the given text into tokens. In fine-tuned models, this process is often optimized for specific contexts or data types. However, when these models encounter numeric values or specific syntax outside their primary training scope, they may misinterpret the context, leading to tokenization problems. For instance, a model might incorrectly tokenize '11' as '111' if it is not adequately attuned to the nuances of RTL code syntax, as exemplified in Listing~\ref{listing-tokenization-problem}. 
Such mismatches in variable sizes and values can result in significant logical errors in the generated code. 
This limitation highlights the need for more context-sensitive tokenization methods in LLMs, especially in specialized domains like RTL code generation, where the accuracy of every token is crucial for the integrity of the output.

\noindent {\bf Inability to capture intended hardware behavior:} 
LLMs often struggle in adhering to specific conventions within RTL coding, which are crucial for preventing undesired hardware behavior. For example, LLMs often overlook the RTL design principle of avoiding the direct assignment of a single element from one buffer to multiple positions in another buffer, favoring the assignment to a register first~\cite{fu2023gpt4aigchip}. Furthermore, the generation of unnecessary ports, as depicted in Listing~\ref{listing-more-ports}, deviates from the best design practices and not only complicate the design but may also lead to conflicts or inefficiencies in hardware implementation. Such deviations from the established RTL coding conventions, underscore the need for more sophisticated understanding and application of hardware design principles in LLMs, ensuring that the generated code not only meets the functional requirements, but also adheres to the practical and efficient design standards essential in RTL development.

\begin{lstlisting}[language={Verilog}, caption={Variable Size Mismatch}, basicstyle=\small, label=listing-tokenization-problem, escapechar=!, float=tp, belowskip=-0.8 \baselineskip]
module instruction_decoder (
    input [31:0] inst,
    output logic [3:0] opcode,
    output logic [1:0] src1,
    output logic [1:0] src2,
    output logic [1:0] dest
);
    assign opcode = inst[3:0];
    assign src1 = inst[7:6];
    assign src2 = inst[9:8];
    !\HighlightA{}!assign dest = inst[111:10];
endmodule
\end{lstlisting}


\begin{lstlisting}[language={Verilog}, caption={More Ports than Necessary}, basicstyle=\small, label=listing-more-ports, escapechar=!]
input logic clk,
input logic rst,
input logic [3:0] opcode,
input logic [15:0] a_in,
input logic [15:0] b_in,
input logic [15:0] scalar_in,
!\HighlightA{}!input logic [$clog2(4)-1:0] vector_wr_addr,
!\HighlightA{}!input logic vector_wr_en,
!\HighlightA{}!input logic [15:0] vector_wr_data,
!\HighlightA{}!input logic [$clog2(4)-1:0] scalar_wr_addr,
!\HighlightA{}!input logic scalar_wr_en,
!\HighlightA{}!input logic [15:0] scalar_wr_data,
!\HighlightA{}!input logic [$clog2(4)-1:0] vector_rd_addr,
!\HighlightA{}!input logic [$clog2(16)-1:0] vector_rd_elem_addr,
!\HighlightA{}!output logic [15:0] result
\end{lstlisting}


\noindent {\bf Improper treatment of user instructions}: Oftentimes, users provide custom directions to language models to guide the process of RTL code generation. Nevertheless, these language models can encounter difficulties when it comes to connecting these instructions in natural language with the correct code generation. For example, when asked to break down a given module into specific sub-modules, they might have trouble accomplishing it~\cite{fu2023gpt4aigchip}. 


The problems described above limits the effective use of LLMs in automated RTL code generation. Some of these problems can be alleviated by employing multiple prompts. In such scenarios, each prompt corresponds to a particular module in the design, and as these modules often have dependencies on one another: the outcomes of earlier prompts need to be transferred to the subsequent prompts. This leads to a lengthy chain of dependency tracking, requiring manual effort, being error-prone, and consuming a significant amount of time. Furthermore, the subsequent prompts in the sequence might not effectively manage lengthy responses from earlier prompts because of their limited context cache size. In this work, however, we focus on automated RTL code generation with a single prompt, which poses event more significant challenges.

An orthogonal approach to address these problems is to refine existing LLMs by fine-tuning them with annotated RTL code. However, this approach encounters challenges due to the limited availability of well-annotated RTL code with accompanying design explanations and the absence of advanced LLMs suitable for fine-tuning. Below, we provide further details on instruction-tuning and fine-tuning of LLMs for RTL code generation. 

\subsection{Instruction-Tuning and fine-tuning LLMs for RTL code generation}

We explore instruction-tuned LLMs specifically for code generation tasks, a methodology that offers distinct advantages over their non-fine-tuned counterparts. These instruction-tuned models demonstrate a greater proficiency in adhering to specified design parameters and enable iterative refinement to correct errors, making them particularly suitable for specialized tasks, such as automated RTL code generation. This instruction-tuning process is crucial for enhancing the precision and utility of the generated code, ensuring that domain-specific nuances and technicalities are accurately captured. However, it is important to note that not all LLMs available for RTL code generation are instruction-tuned. For instance, VeriGen~\cite{thakur2023verigen} represents an LLM developed for Verilog code generation, but it is not instruction-tuned, which is a primary reason for its exclusion from our analysis. Our focus on instruction-tuned models is driven by the need for high accuracy and relevance in generated RTL code, aligning with the specific requirements of the design tasks at hand.

\subsection{Attention Mechanisms in LLMs}
\label{subsec:attention-mechanisms}

In LLMs, the management of attention mechanisms, such as window attention~\cite{beltagy2020longformer} and dense attention pose distinct challenges and trade-offs. Window attention offer efficiency during inference but can increase language modeling perplexity, particularly when the text length exceeds the cache size. This increase in perplexity is mainly due to the exclusion of initial tokens from the focus of the LLM. In contrast, dense attention, when used with a cache size exceeding the given context size, results in significant memory consumption and slower processing speeds. Moreover, it often leads to poorer output quality, as the model is not trained on excessively long sequences of inputs and struggles with extended contexts beyond its training limits.

To mitigate the challenges posed by both window and dense attention mechanisms, an effective approach was introduced, recently, called \textit{attention sink} in the StreamingLLM work~\cite{xiao2023streamingllm}. This \textit{attention sink} concept underscores the crucial role of initial tokens in maintaining the stability of LLM outputs, addressing the instability issues associated with the exclusion of these tokens in window attention techniques. The attention sink approach recognizes the significance of initial tokens, regardless of their distance from the tokens currently being predicted, offering a nuanced understanding and enhancement of attention mechanisms in the LLMs. Furthermore, StreamingLLM refines the approach to token selection and context caching, which are critical to overcoming the limitations of conventional LLMs. By intelligently managing token retention and context caching, StreamingLLM mitigates the common challenges of handling of extended sequences and ensures a consistent and stable attention mechanism while the model operates on sequences of unlimited length. 


In this work, we use \textit{attention sink} mechanism to augment an LLM's ability to preserve specifications during the whole generation of RTL code. Attention sink mechanism addresses the challenges associated with high memory demands and the traditional constraints of LLMs 
handling of long-sequence data. It enables LLMs to maintain critical hardware specifications consistently, thereby enhancing the accuracy and coherence of the generated RTL code.


\section{A Case Study: NPU}

To analyze the RTL code generation capabilities of LLMs with different attention mechanisms (dense, window, and attention sink), we design Neural Processing Unit (NPU) that accelerates bfloat16 vector operations. This design serves as a practical illustration of how these attention mechanisms can be harnessed in a real-world RTL design scenario, thereby providing a tangible comparison of their performance and efficacy in managing complex design tasks.

\begin{figure}[htbp]
\centerline{\includegraphics[width=1\columnwidth]{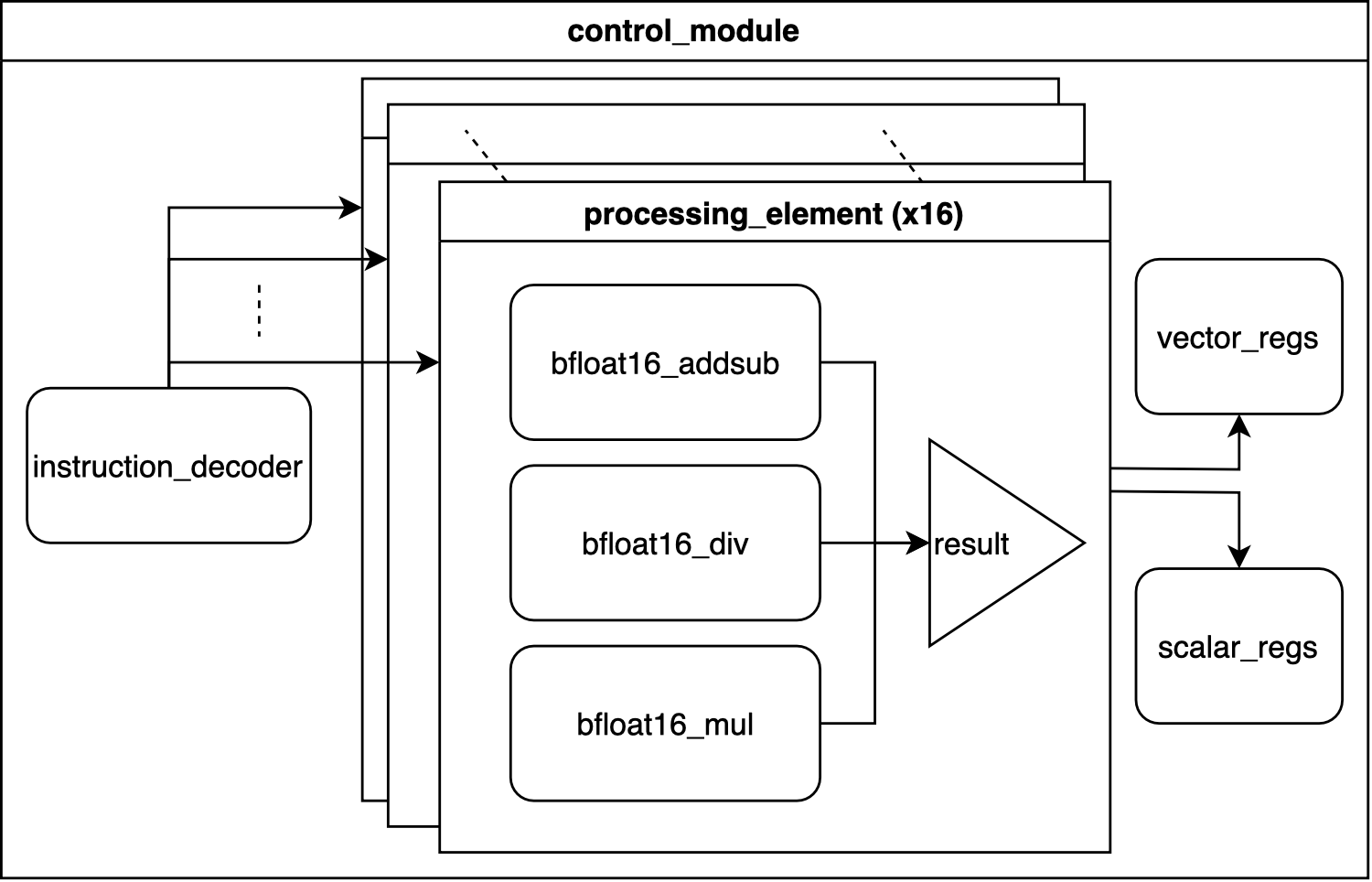}}
\caption{Block diagram of pipelined NPU datapath.}
\label{fig:npu-control}
\vspace{-0.4cm}
\end{figure}

Below is an overview of the critical design choices made for the Neural Processing Unit (NPU), focusing on the adoption of the bfloat16 data type and the implementation of a pipelined architecture (see Fig.~\ref{fig:npu-control}). These choices are pivotal, influencing the performance, efficiency, and capability to manage complex vector operations of NPU
for testing RTL code generation of LLM with different attention mechanisms.
\begin{itemize}
    \item \textbf{Accelerator Design:} We employed an NPU vector coprocessor in our evaluation design to comprehensively explore the capabilities of LLMs in generating code for non-standard architectural designs.
    \item \textbf{Data Type:} We chose bfloat16 as the data type to test the capabilities of LLMs' to work with non standard/less used data types.
    \item \textbf{Pipelined Architecture:} We chose a pipelined design to test the abilities of LLMs' usage of signals across modules.
\end{itemize}

The incorporation of bfloat16 and a pipelined architecture into the NPU design allows testing various attention models, maintaining a balance between efficiency and high performance standards.



\section{Evaluation}
\subsection{Evaluation Setup}


We picked four large language models for evaluation: GPT-4 and the top four models from the BigCode LLM Leaderboard~\cite{BigCode}. The details of these LLMs are given in Table~\ref{tab:llms-used}. Initially, we assessed the ability of these models to accurately follow prompts for converting high-level specifications into RTL code. Subsequently, with the model that best adhered to these instructions, we tested attention sink mechanism against commonly used attention caching techniques in the field. 
Note that we haven't fine-tuned any LLMs specifically for Verilog code generation.  Rather we use existing LLM models with instruction-tuning. This is simply because of the limitations of training large models (scarcity of industry-standard high quality training dataset) and we want to show that it is possible to generate high quality RTL code, even without fine-tuning for Verilog. 

In the evaluation of attention sink enhanced LLMs, we employed a specialized streaming cache (part of the KV cache), tailored specifically for our RTL code generation tasks, using the StreamingLLM framework. This streaming cache is configured to enhance the RTL generation capabilities of the model. Basically, we set the size of the streaming cache to the size of the initial prompt, thereby designating all tokens within the prompt as attention sinks. These tokens, which are critical for retaining the context and specifications of the initial prompt, are permanently kept in memory within the cache. The remainder of the KV cache is dynamically allocated to the newly generated code, effectively combining the static prompt with the dynamic output in a seamless manner. The total KV cache, thus, comprises the streaming cache and the latest generated tokens, filling the context window to optimize both retention of essential context and incorporation of new code. This configuration of the cache ensures that the model efficiently manages extended sequences while maintaining the integrity of both the static and evolving elements of the code generation process. 

The evaluations were performed on the GPU cluster that features NVidia A100 GPUs. The details of the evaluation platform is given in Table~\ref{tab:palamut-cuda-specs}.

\begin{table}[htp]
    \centering
    \caption{List of LLMs Evaluated}
    \label{tab:llms-used}
    \begin{tabular}{|c|c|}
        \hline
        Model Name & Number of Parameters \\
        \hline
        GPT-4 
        \cite{openai2023gpt4} & Not Specified \\
        \hline
        WizardCoder-Python-34B-V1.0
        \cite{luo2023wizardcoder} & 34B \\
        \hline
        Phind-CodeLlama-34B-v2 \cite{Phind} & 34B \\
        \hline
        CodeLlama-34b-Instruct \cite{roziere_code_2023} & 34B \\
        \hline
    \end{tabular}
    \vspace{-0.4cm}
\end{table}

\begin{table}[h]
    \centering
    \caption{Evaluation Platform Specifications}
    \label{tab:palamut-cuda-specs}
    \begin{tabular}{|c|c|}
        \hline
        Component & Specification \\
        \hline
        Server & HP Proliant XL675d Gen10 Plus \\
        \hline
        GPUs & 8x A100 80GB SXM \\
        \hline
        CPUs & 2x AMD EPYC 7742 @ 2.24GHz \\
        \hline
        RAM & 1TB \\
        \hline
    \end{tabular}
\end{table}

We observed distinct differences in how these models responded to our prompts. A notable issue with some of the models was their tendency to rely heavily on comments or non-executable segments, rather than producing functional RTL code. This pattern is exemplified in Listing~\ref{listing-wrong-behavior}, where the model generates extensive comments outlining the intended logic, but falls short in translating these concepts into concrete RTL code. Such outputs, while descriptive, do not fulfill the practical requirement of generating usable RTL code. Among the LLMs being used, WizardCoder able to interpret our prompts more effectively, translating high-level specifications into precise and synthesizable RTL code with greater reliability and less reliance on non-executable textual fillers. This ability to generate functional code with minimal extraneous content made WizardCoder a more suitable candidate for further testing and analysis, particularly in assessing attention sink mechanism against other attention mechanism. Again, the selection of WizardCoder for the rest of the analysis was driven by its demonstrated proficiency in adhering closely to the instructions and generating practical, executable RTL code in response to our prompts.

\begin{lstlisting}[language={Verilog}, caption={Unfunctional Module}, basicstyle=\small, label=listing-wrong-behavior, float=tp, belowskip=-0.8 \baselineskip]
module bfloat16_addsub (
   input logic [15:0] a,
   input logic [15:0] b,
   output logic [15:0] result,
   input is_sub
);
   // Extract sign, exponent, and mantissa
   logic [1:0] sign_a, sign_b, sign_res;
   logic [7:0] exp_a, exp_b, exp_res;
   logic [7:0] mant_a, mant_b, mant_res;
   // ... (extract sign, exponent, and mantissa)

   // Subtract exponents and shift mantissa
   logic [7:0] exp_diff;
   logic [15:0] mant_shifted;
   // ... (calculate exponent difference and shift mantissa)

   // Add or subtract mantissas
   logic [15:0] mant_sum;
   // ... (perform addition or subtraction)

   // Normalization
   logic [15:0] mant_norm;
   // ... (perform normalization)

   // Combine sign, exponent, and mantissa
   result = {sign_res, exp_res, mant_res};
endmodule
\end{lstlisting}

To rigorously evaluate the correctness of the code generated by the selected LLMs, we employed a comprehensive testing methodology focusing on error rates. The error rate is determined by the percentage of code that is generated incorrectly or not generated at all. For cases where an entire module is missing in the generated code, we calculate the error rate by comparing the token size of the correct implementation with the generated output. In cases where all modules are generated, but contain errors, we utilize the DVT IDE, a specialized Verilog compiler, to identify these errors~\cite{dvt-compiler}. Then, we calculate the error rate by assessing the number of tokens that are incorrect or need to be added to make the code functionally correct. This approach allows us to quantitatively measure the correctness  of the code produced by the LLMs, providing a clear metric for comparing their performance and effectiveness in generating RTL code.

\subsection{Results}

\begin{figure}
  \centering
  \includegraphics[width=0.95\columnwidth]{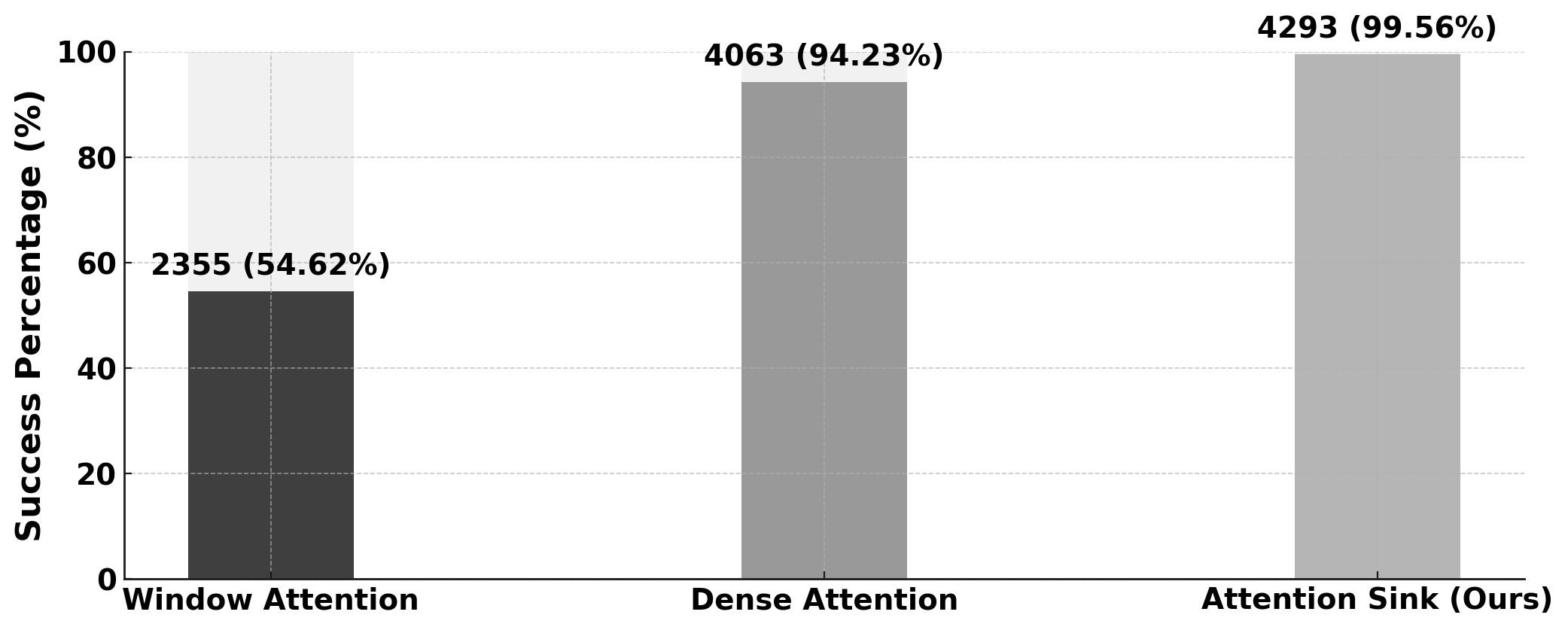}
  \caption{Success percentage of generated tokens with different attention mechanisms using WizardCoder-Python-34B-V1.0. }
  \label{fig:llm-attention-success-rate}
\end{figure}

We tested the quality of code produced for the NPU design by the LLM using dense attention, window attention and attention sink mechanism. The results are shown in Fig.~\ref{fig:llm-attention-success-rate} that demonstrate the superiority of attention sink mechanism over others. The total number of tokens in the correct code is 4312. LLM with attention sink mechanism generates 4293 of 4312 (99.56\%) tokens correctly. Only 16 tokens needed to be fixed for successful compilation of the code for attention sink, whereas dense attention and window attention require 249, and 1957 tokens to be fixed, respectively.

\subsection{Prompt Generation}

We started the evaluation by creating an initial prompt describing the structure of the NPU design and then made gradual improvements based on the errors we observed as the LLM generates RTL code. The final version of the prompt includes high-level specifications, ports for certain modules, such as the top module with wishbone signals (since these can change from one design to another) and basic reminders on maintaining code quality. Listing~\ref{listing-initial-prompt} outlines the content of the initial prompt. We begin the prompt with a very high level specification that indicates the native data type and supported operations.

\begin{lstlisting}[language={}, caption={Initial Prompt}, basicstyle=\small, breaklines=false, label=listing-initial-prompt]
Design and implement these modules for an NPU coprocessor. 
This NPU uses bfloat16 as the native data type. This NPU
accelerates vector-vector and vector-scalar addition, 
subtraction, multiplication, division.
\end{lstlisting}

When ports are not listed in a prompt, LLM tends to add internal signals to the port list. The language model specified the necessary ports like opcode and operators but also outputs the modules' port that it itself initializes that's needed to be controlled in module. To prevent this, we specify the ports on arithmetic modules. This can be seen on some runs within the processing element module that contains bfloat16 arithmetic modules. Port specifications are given to the LLM as shown in Listing~\ref{listing-ports}.

\begin{lstlisting}[language={Verilog}, caption={Prompt given ports}, basicstyle=\small, label=listing-ports]
ports:
 input logic [WIDTH-1:0] a,
 input logic [WIDTH-1:0] b,
 output logic [WIDTH-1:0] result
\end{lstlisting}

Also, every module outlined within the prompt is enhanced with a brief, yet comprehensive explanation that details its intended design and functionality. An example explanation is shown in Listing~\ref{listing-explanation}. These explanations are crucial as they offer a clear overview of each module's role, particularly in relation to the overall architecture of the design. They articulate the purpose of each modules, data handling strategies, and how to interact with other modules, thereby guiding LLMs to generate more precise and contextually relevant RTL code. 


\begin{lstlisting}[language={}, caption={An excerpt from the prompt that explains a module}, basicstyle=\small, breaklines=false, label=listing-explanation]
explanation: instantiates scalar_regs, vector_regs, 
16 processing_elements. Handles load, store, execution 
and writeback.
\end{lstlisting}


Overall, the creation of prompts for RTL code generation followed an iterative approach. Starting with a basic prompt outlining high-level specifications for an NPU design, we continuously refined the prompts based on the output generated by the LLMs. Rather than prescribing specific fixes for the errors, we focused on enriching the prompts with general coding guidelines whenever the model made a mistake. This approach ensured that the LLMs were not just corrected, but were also guided towards a better understanding of the desired coding standards and practices. For instance, when the models added unnecessary internal signals or missed critical functionalities, we did not provide explicit corrections. Instead, we augmented the prompt with more detailed port definitions and operational parameters, along with clear and concise explanations for each module's intended role. This method of iterative enhancement, driven by providing general guidelines rather than specific fixes, allowed LLMs to progressively generate more precise, accurate, and contextually appropriate RTL code. 
Once a prompt includes all necessary detils, LLMs with augmented with attention sink mechanism allow single-shot industry-standard high quality RTL code generation for the desired design. 


\section{Discussion and Future Work}




As mentioned in Section~\ref{subsec:attention-mechanisms}, the role of the attention cache mechanism for code generation is crucial. Dense attention, which stores a complete set of specifications in memory, ensures uninterrupted access to all prompt details. This extensive memory utilization helps in preserving the original context's fidelity, a key aspect in intricate coding tasks. However, there are notable downsides. The extensive data volume not only slows down computation but also may cause the model to repeat or create irrelevant lines of code, particularly in scenarios like signal definitions. Such issues often stem from a weakened context mechanism when handling longer sequences.

On the other hand, the window attention mechanism in LLMs faces challenges in fully capturing the initial specifications in extended sequences, despite its computational efficiency arising from concentrating on smaller input segments. This limitation can cause the model to overlook key elements of the specifications, potentially leading to outputs that don't meet the intended requirements. Thus, while dense attention preserves specifications at the expense of slower computation and a risk of content repetition or inaccuracies, window attention enhances computational speed but risks losing comprehensive context and potentially compromises the accuracy in aligning with the initial prompt.

In contrast to other attention mechanisms, the attention sink mechanism that focuses on the selective retention of initial tokens can theoretically offer a way to mitigate these challenges. By preserving key initial tokens, LLMs can expand the effective context window, potentially enabling more robust in-context learning over longer text sequences. Nevertheless, LLMs that utilize the attention sink mechanism may face challenges when the size of the prompt surpasses the total context window size. Although attention caching is effective in overcoming limitations of context size, if the prompt itself is larger than the available context window, the inability to fully store the prompt in the cache can lead to problems similar to ones shown in Background section. 





As a future work, we plan to investigate on the mechanisms of RTL code generation using LLMs with shorter, more general prompts, especially in handling designs of varying complexities. This approach would challenge the models to interpret and elaborate on less detailed specifications, pushing the boundaries of their inferential capabilities. By experimenting with minimalist prompts, the research can uncover how effectively LLMs can extrapolate intricate hardware design details from limited information, a crucial aspect in evaluating their practical utility in diverse and complex RTL code generation scenarios. This simplified prompt strategy could potentially reveal new insights into the adaptability and innovation of LLMs in architectural exploration and design automation process.

\section{Related Work}

Code generation approaches utilizing LLMs mainly address languages that are well-structured and resemble natural languages or concentrates on simpler tasks like code completion. However, these models are not capable of generating complex, domain-specific hardware designs, which requires well-described specifications and lower-level understanding of modules and their interactions via interfaces. Recently, more studies have been conducted on exploiting LLMs for code generation targeting hardware design.

Hammond et al., conducted the first study on automating the Verilog code generation by fine-tuning an LLM~\cite{hammond2020Dave}. In particular, they explored transfer learning to fine-tune GPT-2 and called their model as DAVE. They also built a tool to generate custom dataset needed for fine-tuning the LLM that exploits several natural language templates that encapsulate different design specification scenarios.

Fu et al., introduced a framework that automates demo-augmented prompt generation pipeline for AI accelerator design~\cite{fu2023gpt4aigchip}. They employ in-context learning to direct LLMs in the process of automating HLS code generation. However, their method still necessitates human intervention for rectifying discrepancies in module interfaces and for merging them to have a code suitable for synthesis. On the other hand, our zero-shot code generation approach produces RTL code with consistent interfaces and reduces the need for human intervention. This is primarily because the modules are generated in response to a single prompt, enabling more effective utilization of the context cache and thereby facilitating the creation of consistent interface implementations.

Thakur et al.,  performed fine-tuning on pre-trained LLMs on Verilog datasets collected from GitHub and Verilog textbooks~\cite{thakur2022benchmarking, thakur2023verigen}. While they created fine-tuned models for RTL code generation, their evaluations demonstrated that the syntactic and functional correctness of the generated code on a large scale design are very poor, making it impractical to automate RTL code generation in real world scenarios.

Chang et al., proposed ChipGPT framework that incorporates a prompt manager before using the LLM model to enable designers to generate high-quality prompts~\cite{chang2023chipgpt}. Notably, ChipGPT generates RTL code without the need for retraining or modifying weights in the  LLMs, making it easily integrable into the latest LLM APIs. However, ChipGPT does not directly modify inaccuracies or complexities in the generated code. Instead, it allows underlying LLM multiple attempts to generate various code versions, refining them iteratively to achieve the final code. Despite aiming to automate RTL code generation using LLMs and reduce manual design efforts, ChipGPT necessitates multiple prompts and manual corrections on the generated code.

Blocklove et al.,  developed a set of benchmarks to evaluate the capabilities of LLMs for functional hardware development and verification~\cite{blocklove2023chipchat}. In particular, they used ChatGPT-4 in interactive mode, to demonstrate a case study that involves a prolonged fully conversational scenario in which 8-bit microprocessor code generated. In their case study, fixing errors in the generated code may require moderate human feedback. If the errors persist, advanced human feedback is provided, involving pinpointing the exact location of the error and specifying the method for correction.

\section{Conclusion}

In this paper, we demonstrated that off-the-shelf LLMs can be used for high quality RTL code generation when these models are enhanced with  \textit{attention sink} mechanism introduced in~\cite{xiao2023streamingllm}. We discuss the details and show how it can help to mitigate the automated RTL-code generation problems, especially for \textit{zero-shot} prompting. Our findings indicate that attention sink and similar mechanisms favorable for future advancements in LLMs, suggesting a trajectory of continued improvement in handling complex, long-sequence tasks in RTL design and beyond.


\bibliographystyle{ieeetr}
\bibliography{references}

\end{document}